\documentclass[lettersize,journal]{IEEEtran}
\usepackage{amsmath,amsfonts}
\usepackage{algorithmic}
\usepackage{algorithm}
\usepackage{array}
\usepackage[caption=false,font=normalsize,labelfont=sf,textfont=sf]{subfig}
\usepackage{textcomp}
\usepackage{stfloats}
\usepackage{url}
\usepackage{verbatim}
\usepackage{graphicx}
\usepackage{cite}
\begin{document}

\title{Microdroplet-Based Communications with \\Frequency Shift Keying Modulation}

\author{Eren Akyol$^{\ast}$, Aysa Azmoudeh$^{\ast}$, Iman Mokari Bolhassan, Pelin Kubra Isgor, and Murat Kuscu
\thanks{*These authors contributed equally.}
        \thanks{The authors are with the Nano/Bio/Physical Information and Communications Laboratory (CALICO Lab) and the Center for neXt-generation Communications (CXC), Koç University, Istanbul, Turkey (e-mail: \{eakyol22, aazmoudeh23, ibolhassan22, pisgor21, mkuscu\}@ku.edu.tr).}
	   \thanks{This work was supported in part by The Scientific and Technological Research Council of Turkey (TUBITAK) under Grant \#120E301, and by the Koç University Seed Research Fund. The authors also acknowledge the use of facilities of the n\textsuperscript{2}STAR-Koç University Nanofabrication and Nanocharacterization Center for Scientific and Technological Advanced Research, and the use of equipment procured through the AXA Research Fund (AXA Chair for Internet of Everything at Koç University - PI: Prof. Ozgur B. Akan).}}

\maketitle

\begin{abstract}

Droplet-based communications has been investigated as a more robust alternative to diffusion-based molecular communications (MC), yet most existing demonstrations employ large ``plug-like" droplets or simple T-junction designs for droplet generation, restricting modulation strategies and achievable data rates. Here, we report a microfluidic communication system that encodes information via the generation rate of sub-$100$ $\mu$m water-in-oil microdroplets using a microfabricated \emph{flow focusing} architecture. By precisely tuning the flow rate of the dispersed-phase (water) via a pressure-regulated flow controller, we implement \emph{frequency shift keying modulation} with four symbols ($4$-FSK). A high-speed optical detection and video processing setup serves as the receiver, tracking system response in the microfluidic channel across different symbol durations ($20$ s and $12$ s) and quantifying error performance. Despite the miniaturized device and channel architecture, our experiments demonstrate programmable and reliable data transmission with minimal symbol errors. Beyond water-in-oil systems, the same encoding principles can be extended to other compartmentalized carriers (e.g., giant unilamellar vesicles, polymersomes) that can also be synthesized via flow focusing techniques, paving the way for biocompatible, robust, and high-capacity communication in intrabody networks and the emerging Internet of Bio-Nano Things.
\end{abstract}

\begin{IEEEkeywords}
Microfluidics, microdroplets, flow focusing, frequency shift keying modulation, microfabrication, molecular communication
\end{IEEEkeywords}

\section{Introduction} 

\IEEEPARstart{I}{nternet} of Bio-Nano Things (IoBNT) envisions seamless networks of engineered micro/nanoscale devices and biological systems, promising for transformative biomedical applications such as intrabody continuous health monitoring and precision drug delivery\cite{kuscu2021internet}. While bio-inspired molecular communications (MC), i.e., information transfer via the release, transport, and detection of biochemical signals, has been extensively studied as a tool to enable this vision, most existing MC frameworks remain fundamentally dependent on diffusion-based strategies\cite{kuscu2019transmitter, jamali2019channel}. These approaches, however, face inherent limitations such as high latency, low throughput, and vulnerability to environmental interference, casting doubt on their practicality in dynamic, complex intrabody environments\cite{kuscu2021fabrication}.

Biological networks, however, do not rely solely on passive diffusion. Instead, they often utilize compartmentalized carriers, such as extracellular vesicles and lipid droplets, to encapsulate molecular cargo, ensuring reliable and robust signal transmission in noisy biological environments\cite{amarasinghe2023cellular,tkach2016communication}. Inspired by these mechanisms, our research experimentally investigates \emph{microdroplets}, i.e., micrometer-scale emulsions, where one fluid is dispersed in another immiscible fluid, as a model system and accessible engineering platform for compartmentalized information delivery in fluidic mediums. 

Droplets have long been employed in lab-on-a-chip systems for applications ranging from cell-free chemical synthesis and drug screening to single-cell assays, owing to their ability to create discrete and isolated microenvironments with precise control over size and composition\cite{guo2012droplet,mazutis2013single}. Building on this versatility, several groups have already studied \emph{droplet-based communications} in lab-on-chip systems, proposing theoretical channel models\cite{galluccio2017capacity}, modulation techniques (e.g., on-off keying, droplet-size encoding \cite{hamidovic2019information, hamidovic2024microfluidic}, speed-based modulation \cite{gallucio2024modeling}), and optical detection systems \cite{bartunik2020colorspecific}. However, existing implementations of droplet-based communications predominantly focus on large ($\gg100$ \textmu m), \emph{plug-like} droplets generated in basic T-junction devices that are often fabricated using 3D-printed molds. These droplets span the entire cross-section of microchannels, restricting their movement to single-file propagation, a configuration incompatible with the envisioned use of vesicle-like, free-floating carriers in complex intrabody environments. Moreover, current systems often lack the resolution to dynamically fine-tune droplet properties, limiting the diversity of modulation strategies that can be implemented, and oversimplifying the challenges associated with droplet propagation and the resulting communication error mechanisms. Crucially, critical design parameters, such as the chemical compatibility of the dispersed/continuous phases and surfactant dynamics, are frequently neglected in these studies, which, therefore, offer limited insight into how engineered droplets (or compartmentalized microstructures broadly) could be optimized as robust, biocompatible information carriers.

In this work, we address these limitations by developing a microdroplet-based communication platform with a custom-designed flow-focusing microfluidic transmitter architecture capable of generating sub-$100$ \textmu m water-in-oil (w/o) droplets with precise hydrodynamic control. Flow-focusing configurations, commonly adopted in lab-on-chip systems for high-throughput synthesis of uniform droplets, provide superior control over droplet properties (e.g., size, generation rate), faster response times, and higher throughput compared to T-junction geometries\cite{anna2003formation,lashkaripour2019performance,chong2016active}. Critically, this approach can also be adapted to produce diverse vesicle-like carriers (e.g., liposomes, polymersomes, or hydrogel droplets)\cite{michelon2017high,petit2016vesicles,deshpande2018chip}, positioning it as a versatile framework for studying compartmentalized communication in intrabody networks.

To encode information, we implement frequency shift keying (FSK) modulation by mapping discrete symbols to distinct droplet generation frequencies at the flow-focusing junction. On the receiver side, we integrate a video-based detection system for real-time measurement of the timing of droplet arrivals at a predefined position in the microchannel, enabling experimental characterization of transmitter performance across a range of symbol durations. We specifically quantify the range and cardinality of stable droplet-generation frequencies, i.e., the set of frequencies at which robust signal formation is maintained, while modulating only the flow rate of the dispersed phase. Communication experiments with $4$-ary FSK modulation show that longer symbol durations yield more stable frequency transitions and lower symbol error rates, whereas shorter durations increase throughput but also introduce higher transition-related errors. These experiments provide a tangible proof-of-concept for microdroplet-based FSK and indicate the scalability of flow-focusing designs to study more advanced communication scenarios.

%The remainder of this paper is organized as follows. In Section \ref{fabrication}, we detail the flow-focusing transmitter design and its soft-lithography microfabrication. Section \ref{experiment} describes the experimental setup and flow, the video-based detection procedure, and the performance metrics for frequency-based droplet modulation. Lastly, Section \ref{conclusion} concludes with a discussion of key findings and future directions in microdroplet-based microfluidic communication.

The remainder of this paper is organized as follows. Section \ref{materialsmethods} details the design and microfabrication of the microfluidic chip and the experimental setup. Section \ref{results} presents the results of the communication experiments with 4-ary FSK with an error analysis under different symbol intervals. Section \ref{conclusion} concludes the paper with key findings and potential future directions.

\begin{figure*}[t]
	\centering
	\includegraphics[width=\textwidth]{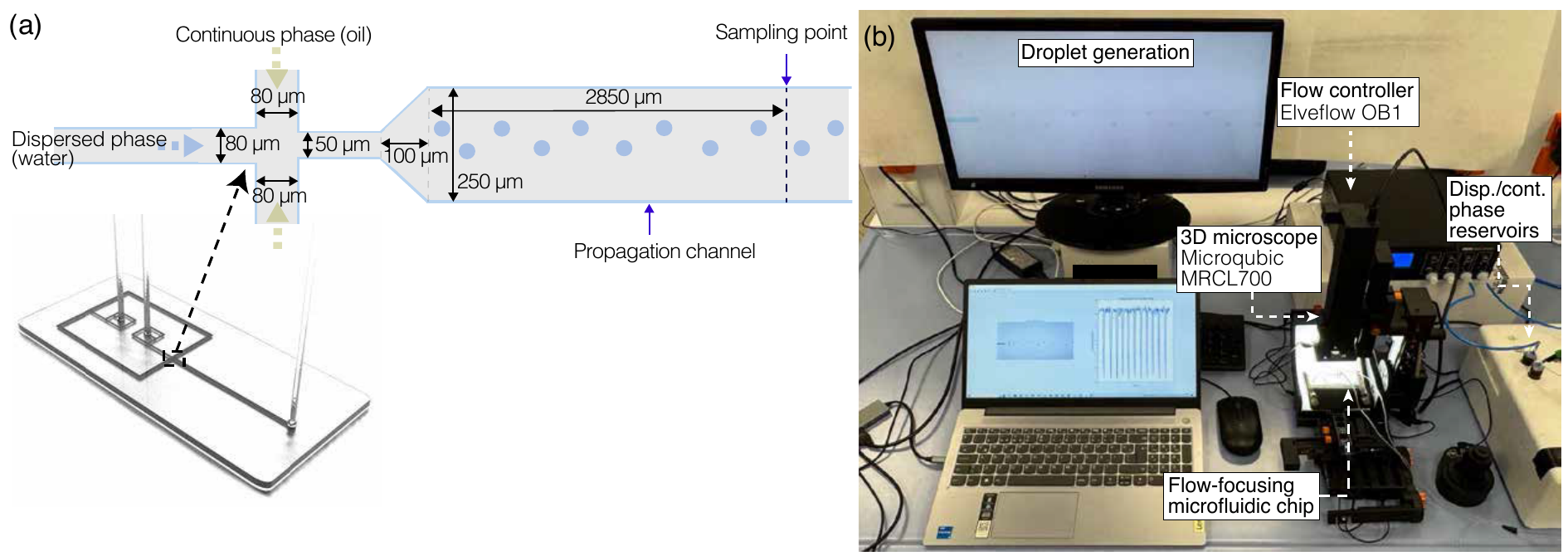}
	\caption{
(a) Top-view layout of the flow-focusing microfluidic chip design, showing the main propagation channel, side channels, and orifice dimensions. \\(b) Experimental setup for microdroplet-based microfluidic communications.}
	\label{fig:layout_setup}
\end{figure*}

\section{Materials and Methods}
\label{materialsmethods}

\subsection{Transmitter Design and Fabrication}
\label{fabrication}

A flow-focusing microfluidic transmitter was designed to achieve uniform sub-$100$ \textmu m droplet formation under precise hydrodynamic conditions. The overall geometry was conceptualized in AutoCAD (Autodesk) and featured distinct channels to manage the dispersed phase (water) and the continuous phase (oil with surfactant). As illustrated in Fig. \ref{fig:layout_setup}(a), the main channel measures $250$ \textmu m in width and transitions to a neck channel ($100$ \textmu m width), which leads to an orifice ($50$ \textmu m width). Two side channels (each $80$ \textmu m wide) introduce the dispersed phase, and droplet formation is driven by the shear forces at the orifice where the continuous phase pinches off the water stream.

The microfluidic chip was fabricated following a typical three-step microfabrication workflow: (i) mold fabrication, (ii) polydimethylsiloxane (PDMS) casting, and (iii) device bonding. \textbf{\textit{Mold fabrication:}} Channel patterns were etched onto a $4$-inch silicon wafer by photolithography and deep reactive ion etching (DRIE, SENTECH SI $500$ C) with $15$ \textmu m etch depth. \textbf{\textit{PDMS casting:}} The Sylgard $184$ elastomer and curing agent (Dow Corning) were mixed at a $10:1$ ratio (w/w), then degassed under vacuum to remove air bubbles. Before pouring the mixture, the etched silicon mold was treated with HMDS vapor to improve PDMS separation from the silicon mold, then the mixture was poured over the etched silicon mold to ensure accurate replication of the channel features. Thermal curing at $70$°C resulted in solidified PDMS with embedded negative replicas of the flow-focusing design. \textbf{\textit{Device bonding:}} Cured PDMS was peeled off the mold, and inlet/outlet ports ($600$ \textmu m diameter) were punched using precision biopsy punches. Both the PDMS layer and a glass substrate were thoroughly cleaned and treated with oxygen plasma to promote covalent bonding. The plasma-treated surfaces were then brought into contact, forming a robust, leak-free seal around the microfluidic channels.

\subsection{Experimental Setup}
\label{experiment}
A microdroplet-based communication system was assembled by coupling the fabricated flow-focusing transmitter to a pressure-regulated flow controller (OB$1$ MK$3$, Elveflow), as shown in Fig. \ref{fig:layout_setup}(b). Two fluid reservoirs provided the continuous and dispersed phases: mineral oil (Sigma-Aldrich, M$8410$) containing $1\,\%$ (w/w) Span~$80$ (Sigma-Aldrich, $85548$), and deionized (DI) water, tinted with a commercially available food dye for imaging contrast. The Span~$80$ surfactant was chosen to stabilize the water-in-oil emulsion, reducing coalescence and maintaining uniform droplet formation. 

\noindent \textbf{Flow Conditions and Imaging:} The continuous-phase pressure, $P_{\mathrm{cont}}$, was held at $1200\,\mathrm{mbar}$, while the dispersed-phase pressure, $P_{\mathrm{disp}}$, was varied to adjust the droplet generation frequency. Microdroplets were monitored using a $3$D microscope (Microqubic MRCL$700$) in a $2$D inverted configuration, capturing real-time videos of the w/o droplets as they were formed and propagated downstream. 

\noindent \textbf{Experimental Flow:} The experiments proceeded in two main phases: frequency modulation characterization and communication performance testing. First, we measured both the generation frequency and the arrival frequency of droplets at a predefined sampling point (serving as the receiver position) along the propagation channel which is located $2850$ \textmu m away from the neck channel, by systematically varying $P_{\mathrm{disp}}$ while keeping $P_{\mathrm{cont}}$ fixed. From these measurements, we computed the mean and variance of the generation and arrival frequencies to identify appropriate boundaries for a \mbox{4-ary} FSK scheme. In the subsequent communication phase, the transmitter alternated among selected four $P_{\mathrm{disp}}$ levels, each mapped to a distinct symbol. A video-based receiver then classified the droplet arrival frequencies at the receiver position using predefined thresholds, enabling quantification of symbol detection accuracy and overall error performance. 

%\noindent \textbf{Data Collection and Analysis:} At each $P_{\mathrm{disp}}$ setting, approximately $20$ droplets were produced, and the time intervals between consecutive droplet generations were recorded at the beginning of the propagation channel, as depicted in Fig. \ref{fig:layout_setup}(a). From these measurements, we computed mean and variance of the droplet intervals, which formed the basis for defining the droplet-generation frequencies used in the design and analysis of FSK modulation. Subsequently, symbol thresholds were identified to partition the frequency range into distinct encoding bins.

\begin{figure*}[t]
	\centering
	\includegraphics[width=0.97\textwidth]{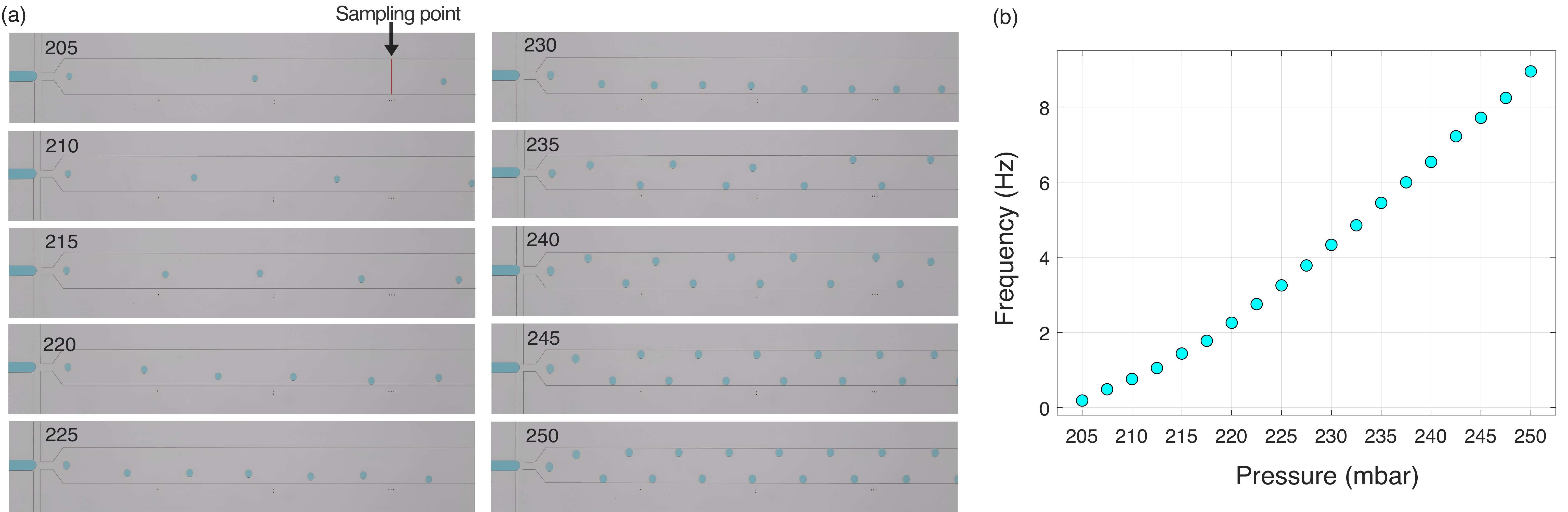}
	\caption{(a) Visualization of sub-$100$ \textmu m w/o droplet formation at various dispersed-phase pressures (applied at the corresponding inlet) in the flow-focusing microfluidic channel. The water phase is tinted with a commercial dye for clarity, and pressures (in mbar) decrease from top to bottom. 
	(b) Measured relationship between dispersed-phase pressure and droplet generation frequency, illustrating a monotonic increase in droplet frequency with increasing pressure.}
	\label{Fig:frequencyrange}
\end{figure*}

\begin{figure}[b!]
	\centering
\includegraphics[width=1\columnwidth]{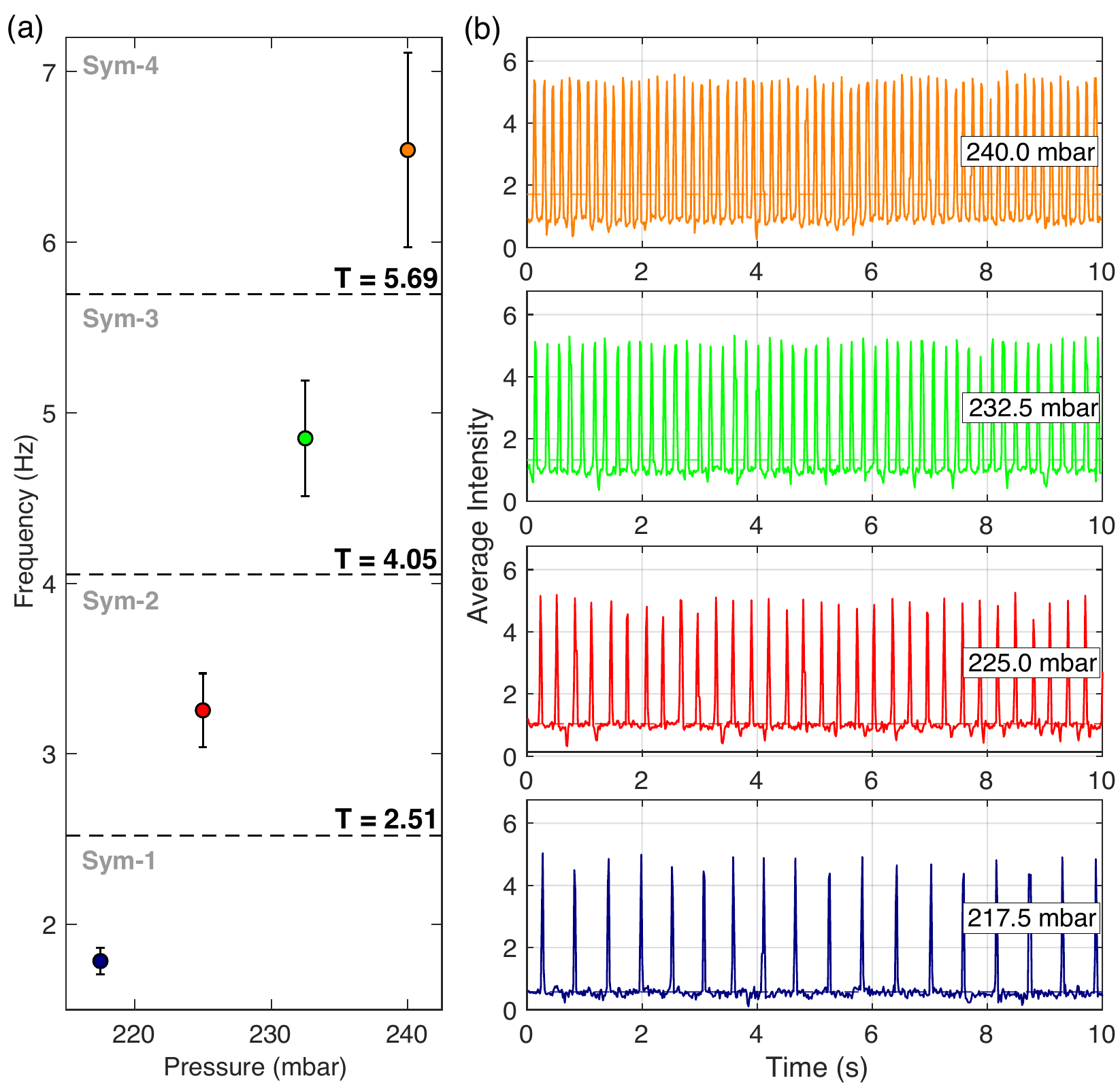}
	\caption{(a) Thresholding of droplet arrival frequencies at selected dispersed-phase pressures. The dashed lines indicate the frequency boundaries for mapping each pressure setting to a unique symbol, Sym-\{1,2,3,4\}. (b) Time-domain intensity plots of droplet arrival at four different pressures.}
	\label{Fig:thresholds}
\end{figure}

\section{Results and Discussion}
\label{results}

\subsection{Characterization of Frequency Modulation}

We first characterized the range and resolution of the droplet generation frequencies by varying $P_{\mathrm{disp}}$ from $205\,\mathrm{mbar}$ to $250\,\mathrm{mbar}$, while maintaining $P_{\mathrm{cont}}$ at $1200\,\mathrm{mbar}$. Each $P_{\mathrm{disp}}$ level was held constant until $20$ droplets were produced, and the time intervals between consecutive droplet generations were recorded at the beginning of the propagation channel (Fig. \ref{fig:layout_setup}(a)). Even small adjustments in $P_{\mathrm{disp}}$ yielded clearly distinguishable droplet generation frequencies, as evidenced by visual inspection in Fig. \ref{Fig:frequencyrange}(a) and the monotonic frequency increase in Fig. \ref{Fig:frequencyrange}(b). Moreover, increasing $P_{\mathrm{disp}}$ raised the generation frequency consistently, indicating the robustness and programmability of the flow-focusing transmitter for the implementation of diverse communication scenarios.

%From these measurements, we computed mean and variance of the droplet intervals, which formed the basis for defining the droplet-generation frequencies used in the design and analysis of FSK modulation.

\begin{figure*}[t!]
	\centering
\includegraphics[width=0.8\textwidth]{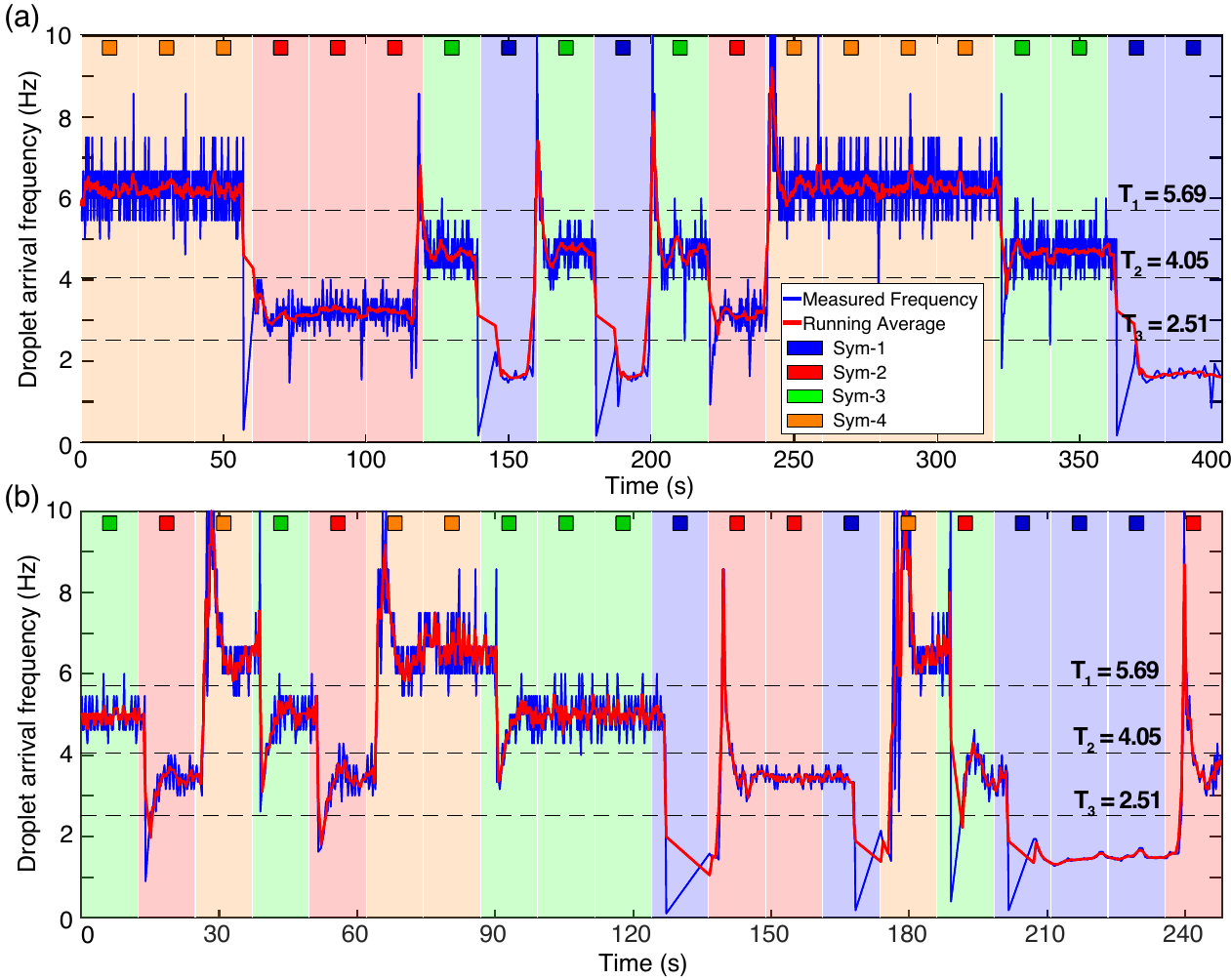}
	\caption{Receiver response to randomly generated $20$-symbol sequence with intervals of (a) $20$ s and (b) $12$ s, where the background colors represent the transmitted symbols, and the small squares denote decoded symbols.} 
	\label{Fig4:comm}
\end{figure*}

Building on these observations, we identified four distinct generation frequencies to serve as symbols in the FSK scheme and then characterized the corresponding arrival frequencies at thesampling point (see Fig.~\ref{Fig:thresholds}). Specifically, for each $P_{\mathrm{disp}}$ setpoint, we recorded the inter-arrival durations of 20 consecutive droplets, converted these durations into arrival frequencies, and calculated their mean and variance. We then set frequency thresholds by taking the midpoint between consecutive mean frequencies, thereby partitioning the measured frequency range into four distinct bins. Figure~\ref{Fig:thresholds}(a) shows these thresholds ($T_1 = 2.51\,\mathrm{Hz}$, $T_2 = 4.05\,\mathrm{Hz}$, $T_3 = 5.69\,\mathrm{Hz}$) alongside the mean frequencies obtained at $P_{\mathrm{disp}} \in \{217.5, 225, 232.5, 240\}\,\mathrm{mbar}$. The corresponding time-domain intensity plots at the sampling point, shown in Fig.~\ref{Fig:thresholds}(b), capture the real-time pixel intensity spikes caused by each passing droplet. Each $P_{\mathrm{disp}}$ level clearly produces a distinct and stable frequency, displaying minimal overlap between symbols.

\subsection{Communication Experiments}

Once the symbols and detection thresholds were determined, we conducted communication experiments to evaluate the end-to-end performance of the system, using a 4-ary FSK modulation scheme. During these experiments, random symbol sequences were generated, with the flow-focusing transmitter switching between selected four symbols by adjusting $P_{\mathrm{disp}}$ accordingly and video processing pipeline served as the receiver. After that, the measured frequencies, thresholds, and symbol classifications were analyzed based on the smoothed frequencies at the midpoint of the detection windows and the results for the first $20$ transmissions with 20\,s and 12\,s  symbol intervals are presented in Fig. \ref{Fig4:comm}. This setting enabled a direct investigation of the trade-off between higher data throughput (shorter symbol interval) and improved decoding reliability (longer symbol interval). 

% discuss 20 seconds
When the symbol duration was set to $20\,\mathrm{s}$, we observed consistent, error-free classification of all $20$ transmitted symbols, as plotted in Fig.~\ref{Fig4:comm}(a). The extended symbol duration stabilized the droplet generation frequency before the midpoint of the detection windows by allowing more droplets to be generated and a greater number of droplet intervals to be measured at the sampling point. Although minor timing offsets arose in the detection windows due to pressure-regulated flow controller's latency when switching $P_{\mathrm{disp}}$, the $20$\,s symbol interval effectively compensated for these small shifts, maintaining accurate symbol classification throughout.

% discuss 12 seconds and the reasons for the error
We next reduced the symbol interval to $12$\,s to evaluate the end-to-end communication reliability under higher data transmission rates. In this scenario, the droplet generation frequency had less time to stabilize, and fewer droplet arrivals were available within a detection window for classification. Moreover, delays in the flow controller became more pronounced, occasionally shifting the frequency transition outside the nominal $12$ s windows. To synchronize the detection windows with these delayed transmissions for consistent, accurate classification, we slightly extended the detection window to $12.4$\,s. As a result, despite the limited symbol duration and the use of very simple detection method, we observed only \emph{$1$} classification error within the first $20$ transmissions. Moreover, this classification errors for the shorter symbol durations, can be easily mitigated through alternative approaches such as performing the classification near the end of each interval or implementing more sophisticated detection algorithms. Overall, these experiments confirm the end-to-end robustness and programmability of the microdroplet-based communication system with an FSK modulation scheme.

\section{Conclusion}
\label{conclusion}
%In this study, we introduced a microdroplet-based communication system platform consisting of a flow focusing-based transmitter and video-based detection system as the receiver. By modulating the pressure of the dispersed phase, we successfully encoded information through droplet generation frequencies, implementing a reliable $4$-symbol FSK modulation scheme. Our results demonstrated that higher pressures led to faster droplet generations, indicating the programmability of the transmitter for encoding and transmitting information. Overall, the system was able to transmit symbols with high accuracy, and low error rate, making this approach promising for biomedical applications such as intrabody networks and the IoBNT. Future improvements will focus on integrating channel coding, more advanced detection algorithms, and biocompatible carriers including liposomes or polymer vesicles to improve reliability and diversify the communication scenarios. Moreover, incorporating feedback-controlled droplet generation and machine learning-based detection could further optimize the system performance. With these advancements, droplet-based communication approach has the potential to become a practical and robust method for biological communication systems.

In this work, we introduced a flow-focusing microfluidic platform for microdroplet-based communication, successfully implementing a 4-ary FSK modulation scheme via precise control of the dispersed-phase pressure. We showed that by varying this pressure, droplet generation frequency can be tuned over a large range with well-defined and distinguishable frequency levels, enabling reliable symbol encoding. A video-based detection system, acting as the receiver, accurately captured droplet arrival frequencies in real time, achieving near-error-free symbol recognition over varying symbol durations (20 s and 12 s). Our findings highlight the importance of system stabilization time in minimizing transition errors: longer symbol durations increased accuracy, while shorter intervals improved throughput but introduced occasional classification errors. These results demonstrate the versatility and potential of microdroplet-based communication, particularly for biomedical applications in intrabody networks and the IoBNT, where robust compartmentalized carriers are essential. Future work will focus on developing channel coding and detection techniques, and more biocompatible carrier formulations (e.g., liposomes, polymersomes) to further improve data rates, reliability, and applicability in biomedical contexts.

% \section{Acknowledgments}, should be on the first page.

\bibliographystyle{IEEEtran}
\bibliography{manuscript}

\vfill

\end{document}